\documentclass[%
reprint,
amsmath,amssymb,
aps,
longbibliography
]{revtex4-2}

\usepackage{graphicx}
\usepackage{dcolumn}
\usepackage{bm}
\usepackage{hyperref}
\usepackage[dvipsnames]{xcolor}

\newcommand{\pd}[2]{\frac{\partial #1}{\partial #2}}
\newcommand{\rv}[1]{\textcolor{black}{#1}}

\begin{document}

\preprint{APS/123-QED}

\title{Intrinsic threshold electric field for domain wall motion in ferroelectrics \\ based on discretized phase-field model}

\author{Huanhuan Tian}
\affiliation{%
 Zhangjiang Laboratory, Shanghai, China
}%
\author{Jianguo Yang}
\email{yangjg@zjlab.ac.cn}
\affiliation{%
 Zhangjiang Laboratory, Shanghai, China
}%
\author{Ming Liu}
\email{liuming@fudan.edu.cn}
\affiliation{%
Frontier Institute of Chip and System, Fudan University, Shanghai, China
}%
\affiliation{%
 Institute of Microelectronics, Chinese Academy of Sciences, Beijing, China. 
}

\date{\today}

\begin{abstract}
With the development of ferroelectric memories, it is becoming increasingly important to understand the ferroelectric switching behaviors at small applied electric fields.  In this \rv{paper}, we use discretized phase-field models to  systematically investigate the intrinsic threshold electric field (TEF) to drive flat 180$^\circ$ and 90$^\circ$ domain walls (DWs), \rv{which can not be captured by continuum models. The results} show that \rv{this TEF increases as the ratio of DW width to unit cell size decreases, and it becomes significant if the DW width is thinner than two unit cells. The} results are qualitatively consistent with existing first-principles studies and \rv{cryogenic} experiments. In addition, this work \rv{proposes a conceptual model to explain the activation electric field (AEF) observed in experiments at room temperature}. This work improves the understanding of DW motion kinetics at small applied fields, and shows that the mesh size and orientation are both important for the phase-field modeling of the above process.


\end{abstract}

\maketitle

\rv{\section{Introduction}}

In the last decade, ferroelectric memories have attracted extensive attention from the industry and academia \cite{schroeder_fundamentals_2022, ramaswamy_nvdram_2023, silva_roadmap_2023}, since the discovery of the new CMOS-compatible and scalable ferroelectric material, hafnia ferroelectrics, in 2011 \cite{boscke_ferroelectricity_2011}. Innovations to improve the performance of ferroelectric memories include coercive field reduction \cite{wang_stable_2023},  non-destructive read \cite{deng_comparative_2023}, and high-density 3D architectures  \cite{deng_comparative_2023, fu_first_2023}, which all require in-depth understanding of ferroelectric switching especially at small applied electric fields. 

Domain walls (DWs), which separate domains with different polarization directions but the same parent phase orientation, are commonly considered critical for ferroelectric switching \cite{tagantsev_domains_2010}. 
\rv{For example, DWs separating domains with antiparallel (perpendicular) spontaneous polarization are called 180$^\circ$ (90$^\circ$) DWs.  Other angles, such as 71$^\circ$, 109$^\circ$, 60$^\circ$, 120$^\circ$ DWs, are also possible \cite{Marton_domain_2010, tagantsev_domains_2010}.}
The ferroelectric switching is generally believed to be controlled by two processes: the nucleation of reverse domains (the formation of DWs),  and the DW motion \cite{ishibashi_note_1971, tagantsev_non-kolmogorov-avrami_2002}. Electrical measurements cannot directly separate the two processes, and can only provide an effective DW \rv{velocity} $v_{DW}$ \rv{(or, switching speed)}, which is usually found to follow the Merz law \cite{merz_switching_1956, pulvari_phenomenological_1958, jo_nonlinear_2009} at room temperature:
\begin{equation}
    v_{DW} \sim \exp(-(E_a/E_{app})^s),
    \label{eq:Merz}
\end{equation}
where $E_a$ is the activation electric field (AEF) for DW motion, $E_{app}$ is the applied electric field\rv{, and the index $s$ is usually around 1}. Typically $E_a$ is much larger than the coercive field $E_c$ \cite{merz_domain_1954, lee_domains_2021, alessandri_switching_2018}.  At cryogenic temperature, a sharp corner on $v_{DW}$-$E$ curve has been observed in electrical measurements \cite{jo_nonlinear_2009}: 
\begin{equation}
    v_{DW}  \sim [\mathcal{R}(E_{app}-E_{th})]^m,
    \label{eq:Eth}
\end{equation}
where $E_{th}$ is the threshold electric field (TEF) to drive DWs, $\mathcal{R}(:)$ is the ramp function, and the index $m$ is typically around 1. Apparently, $E_{th}$ should be no larger than $E_c$ \cite{jo_nonlinear_2009, liu_intrinsic_2016}.

Different theories have been used to explain AEF and TEF for DW motion. Note that the continuum phase-field theory of a flat DW leads to $E_{app}$-independent DW mobility without AEF or TEF \cite{su_continuum_2007}.  AEF has been widely explained by the thermally-activated nucleation of a triangle, square, or circular protrusion on existing flat 180$^\circ$ DWs \cite{miller_mechanism_1960,  shin_nucleation_2007, liu_intrinsic_2016}. However, this thermal nucleation theory predicts $E_a\sim 1/T$, leading to infinitely large $E_a$ near cryogenic temperature, which contradicts the experimental observation. In addition, the assumption of protrusions on flat DWs seems to be artificial, which cannot be automatically reproduced in first-principles calculations and phase-field simulations. Indeed, first-principles calculations for ideally-flat DWs can also give an energy barrier for DW motion \cite{liu_intrinsic_2016, choe_unexpectedly_2021, lee_scale-free_2020, ding_atomic-scale_2020}, but in a TEF way. Such a barrier has been qualitatively explained by John W. Cahn using a discretized phase-field model \cite{cahn_theory_1960}. Though very inspiring, Cahn's work is limited to special cases (thick, flat 180$^\circ$ DW), has errors in derivations (discussed later in this work), and is underestimated in the community possibly due to the gap to explain the AEF observed in experiments.


In this \rv{paper}, we systematically investigate the intrinsic TEF \rv{to drive 180$^\circ$ and 90$^\circ$ DWs based on discretized phase-field models, for general second-order phase transition without elastic energy (Sec.II), and a real ferroelectric material BaTiO$_3$ (Sec.III).  We show that this TEF decreases as the ratio of DW width to mesh size (or unit cell size, if we assume the numerical mesh coincides with unit cells in crystals) increases, supported by analytical solutions, numerical results, and scaling laws. Finally, we discuss on the significance of this work, and propose a possible, new explanation of the AEF phenomena at room temperature in Sec.IV.} 


\rv{\section{Model and results for general second-order phase transition without elastic energy}}
\rv{In this section, we limit the discussion to second-order phase transition of ferroelectrics without elastic energy. Therefore, the model involves only a few parameters ($a$, $b$, $\eta$, $\kappa$, $\kappa'$, $\tau$, $\epsilon_b$, as explained later), making it easy to derive analytical solutions, explore the full parameter space, and grasp the scaling law and key qualitative conclusions. We leave the discussion of elastic energy and real ferroelectric materials to Sec.III. } 
\rv{\subsection{The model}}
\rv{In this section, we write the total free energy of the ferroelectrics as} \cite{wang_understanding_2019, hlinka_phenomenological_2006}:
\begin{equation}
    F = \int{ (f_{landau} + f_{grad} + f_{elec}) }dV,
\end{equation}
where the Landau energy $f_{landau} = \frac{1}{2} a (P_x^2 + P_y^2)+ \frac{1}{4} b (P_x^4 + P_y^4) + \frac{1}{2}\eta P_x^2 P_y^2$, the gradient energy \rv{$f_{grad} = \frac{1}{2} \kappa (|\pd{P_x}{y}|^2 + |\pd {P_y}{x}|^2) + \frac{1}{2}\kappa' (|\pd{P_x}{x}|^2 + |\pd {P_y}{y}|^2)$}, and the electrostatic energy $f_{elec} = -\mathbf{P}\cdot \mathbf{E} - \frac{1}{2}\epsilon_0 \epsilon_{r,b} |\mathbf{E}|^2 $. Here $P_x$ and $P_y$ are the spontaneous polarization in $x$ and $y$ direction ($\mathbf{P} = [P_x, P_y]$), $\mathbf{E} = -\nabla \phi$ is the electric field, $\epsilon_0$ is the vacuum permittivity, $\epsilon_{r,b}$ is the background relative permittivity,  $a$, $b$, $\eta$ are Landau parameters, and $\kappa$\rv{, $\kappa'$ are} the gradient penalty parameter\rv{s}. Then the dynamics of ferroelectric switching can be described by  \cite{wang_understanding_2019}
\begin{equation}
    \frac{\tau}{\epsilon_0}\pd{\mathbf{P}}{t} = -\frac{\delta F}{\delta \mathbf{P}}  \text{ and }  \nabla\cdot (\mathbf{P} + \epsilon_0 \epsilon_{r,b} \mathbf{E}) = 0
    \label{eq:pde}
\end{equation}
in combination with proper boundary conditions, where $\tau$ is a time constant. We define the thickness of ferroelectric film as $h$. We assume that the energy minimum of $f_{landau}$ occurs at $[0, \pm P_0]$, $[\pm P_0, 0]$, and take into account other grain orientations by coordinate transformations. In addition, we assume that $E_{app}$ is positive if the applied field \rv{points down the} $y$ axis.

We can derive the following intrinsic relations based on continuum theory. Note that $\eta \neq 3b$ leads to anisotropic permittivity, charged DWs, and absence of analytical solutions, for 90$^\circ$ switching \cite{xiao_depletion_2005}.
\rv{And the following switching electric field is the effective component along DWs or $-\Delta \mathbf{P}$.}
(1) Basic properties: the remnant polarization $P_0 = \sqrt{-b/a}$, the longitudinal dielectric constant  (parallel to $\mathbf{P}$) $ \epsilon_{r,\parallel} =  \epsilon_{r,b} + \frac{1}{ \epsilon_0 (a + 3b P_0^2)} $ and the transverse dielectric constant (perpendicular to  $\mathbf{P}$) $\epsilon_{r,\perp} =   \epsilon_{r,b} + \frac{1}{\epsilon_0 (a + \eta P_0^2)}$. (2) Bulk switching properties ($\eta = 3b$):  the energy barrier $g_0 = \frac{1}{4} b P_0^4$ and TEF $E_0 = \frac{2}{3}\sqrt{\frac{1}{3}}b P_0^3$ for uniform 180$^\circ$ switching, the energy barrier $g_0/2$ and TEF $\frac{1}{\sqrt{2}}E_0$ for uniform 90$^\circ$ switching.  (3) \rv{180$^\circ$ }DW properties \rv{\footnote{The analytical solution is $P_y = -P_0 + 2P_0 \tanh(\frac{x-x_{DW}}{\delta})$, $P_x = 0$}}: the DW characteristic width $\delta =\frac{1}{P_0}\sqrt{\frac{2\kappa}{b}} = \frac{3\gamma_0}{8g_0}$ (around half real DW width), the  energy $\gamma_0 = \frac{P_0^3}{3}\sqrt{8\kappa b}$ and mobility (\rv{defined as} DW normal velocity $v_{DW}$ divided by \rv{the electric field along DWs}) $\mu_0 =\frac{3\epsilon_0 \delta}{2\tau P_0}$ \cite{collins_dynamics_1979, tagantsev_domains_2010}.  \rv{(4) 90$^\circ$ DW properties ($\eta = 3b$) \rv{\footnote{The analytical solution is $P_y = -\frac{P_0}{\sqrt{2}} + \sqrt{2}P_0 \tanh(\frac{x-x_{DW}}{\delta}\sqrt{\frac{2\kappa}{\kappa' + \kappa}})$ , $P_x = P_0/\sqrt{2}$}}: DW width $\delta \sqrt{\frac{\kappa' + \kappa}{2\kappa}}$ (which equals $\delta$ if $\kappa' = \kappa$), the energy $\gamma_0/2$ and mobility $\sqrt{2}\mu_0$}. 

\begin{figure}[t]
\centering
\includegraphics[width = 8.6cm]{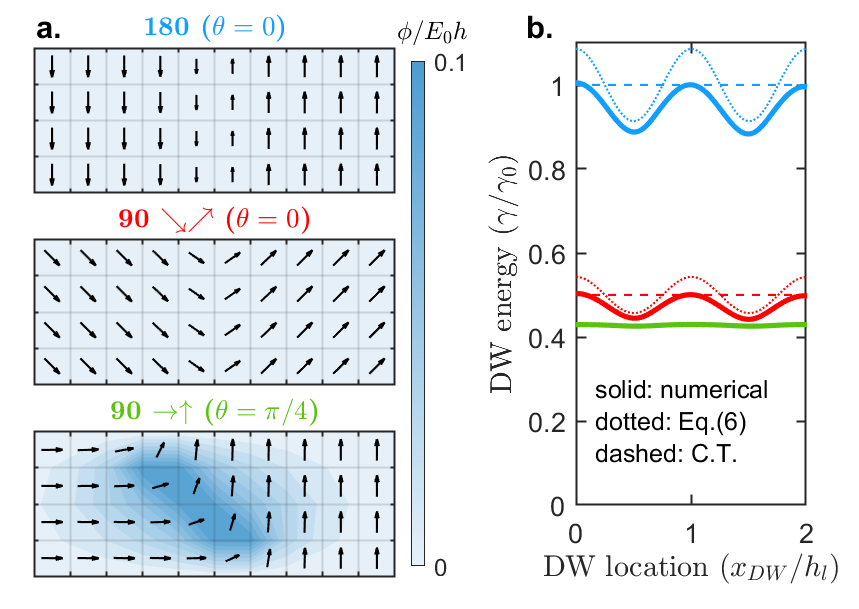}
\caption{\label{fig:figure1} (a) The equilibrium local polarization and electric potential, and (b) the \rv{energy landscape as DW proceeds},  for three types of DWs with \rv{$\delta/h_l = 0.8$, $\eta = 3b$, $\epsilon_{r,b}=\epsilon_{r}/6$, $\kappa' = \kappa$}. Electrodes (constant potentials) are placed on top and bottom boundaries. $\theta$ is to describe the DW orientation. }
\end{figure}

\rv{\subsection{Basic results for DW motion}}
Next, we turn to the discretized model. We assume \rv{the mesh and unit cell have the same size $h_l$}, and apply the finite volume method with staggered mesh for $\phi$ and $\mathbf{P}$ \rv{to get the numerical results}.  \rv{We study three types of DWs: one 180$^\circ$ DW, and two 90$^\circ$ DWs (``90 $\searrow \nearrow$" and ``90 $\rightarrow\uparrow$"),  as shown in Figure \ref{fig:figure1}(a).  We let the remnant dipole in each unit cell to be parallel to cell sides for  180$^\circ$ DW and ``90 $\rightarrow\uparrow$" DW, but to cell diagonals for ``90 $\searrow \nearrow$" DW. Note that the ``90 $\searrow \nearrow$" DW is not commonly seen in popular ferroelectric materials under normal conditions, but is also possible \cite{Marton_domain_2010}. In this part, we assume $\kappa' = \kappa$, $\eta = 3b$, $h = 8h_l$ ($h =4h_l$ only for  Figure \ref{fig:figure1}(a) and Figure \ref{fig:figure3}(d)), $\epsilon_{r,b} = \epsilon_r/6$ for all the simulations.}

Figure \ref{fig:figure1} shows typical numerical solutions of the discretized model for the three DWs (given $\delta/h_l = 0.8$, $\eta = 3b$, $\epsilon_{r,b}=\epsilon_{r}/6$, \rv{$\kappa' = \kappa$}). As shown in Figure \ref{fig:figure1}(a), the polarization is reduced near the DWs, where the switching should be easier compared to bulk domains. In addition,  the ``90 $\rightarrow\uparrow$" DW perturbs the electric field while the 180 DW and the ``90 $\searrow \nearrow$" DW do not. The reason is that the bound charge of adjacent cells cannot be fully compensated in ``90 $\rightarrow\uparrow$" DW.  Figure \ref{fig:figure1}(b) shows the energy landscape for DW motion, extracted from the time evolution of total energy as DW proceeds at some applied field $E_{app}$ (driven force has been subtracted). As expected, the total energy along with DW location has a period of $h_l$, due to translational symmetry, as also seen in first-principles studies \cite{lee_scale-free_2020, choe_unexpectedly_2021}. Since we put each dipole at the center of each cell, the energy peaks and valleys occur when DW passes the cell centers and edges, respectively. As a contrast, the continuum theory (C.T.) gives zero energy barrier since the total energy does not change with DW location. Furthermore, Figure \ref{fig:figure1}(b) shows that  the 180 DW has the highest DW motion barrier, followed by the  ``90 $\searrow \nearrow$" DW, and then the ``90 $\rightarrow\uparrow$" DW.

\begin{figure}[t]
    \centering
    \includegraphics[width= 8cm]{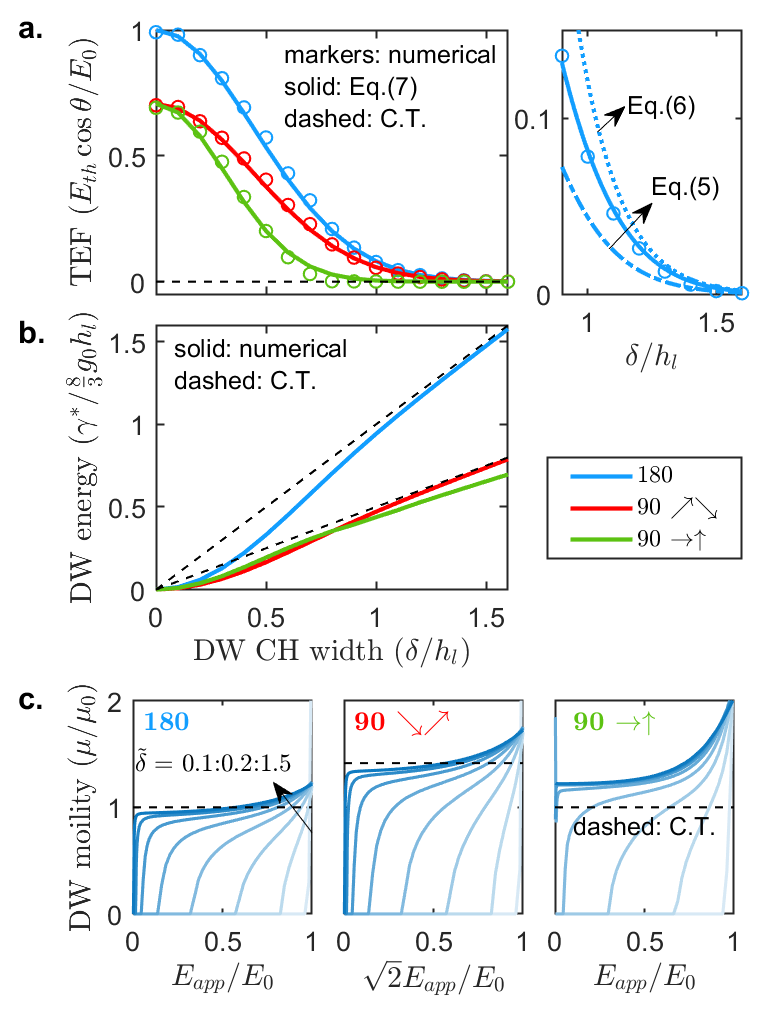}
    \caption{(a) The threshold electric field (TEF), (b) DW energy, and (c) DW mobility for different DW characteristic (CH) length, based on the discretized model with \rv{$\eta = 3b$, $\epsilon_{r,b}=\epsilon_{r}/6$, $\kappa' = \kappa$, }$h=8h_l$. }
    \label{fig:figure2}
\end{figure}

Cahn was the first to seek to analytically derive the periodic picture of $\gamma(x_{DW})$ for 180$^\circ$ DW without \rv{thermal} nucleation theory \cite{cahn_theory_1960}. He substituted the continuum solution $P = -P_0 + 2 P_0 \tanh(\frac{x-x_{DW}}{\delta})$ into a discretized form of total energy $F = \sum_n \frac{1}{4}b(P_0^4 - P_n^4)$ \footnote{The key step of this derivation is $\sum_n f_{grad,n} = \frac{\kappa}{2 h_l^2} \sum_n (P_{n+1} - P_n)^2 =  \frac{\kappa}{2 h_l^2} \sum_n P_n(2P_n - P_{n-1} - P_{n+1}) =   \frac{1}{2} \sum_n P_n (\frac{\partial f_{landau}}{\partial P})_n$} (where $n$ is cell label) and then used Poisson summation to obtain:
\begin{equation}
    \gamma_{180,ana1} = \gamma_0 [1 + 2\pi^2 \tilde{\delta} (1-\frac{\pi^2 \tilde{\delta}^2}{2}) \mathrm{csch}(\pi^2\tilde{\delta}) \cos (2\pi \tilde{x}_{DW})  ],
    \label{eq:gamma_cahn}
\end{equation}
where $\tilde{\delta} = \delta/h_l$, $\tilde{x}_{DW} = x_{DW}/h_l$, and higher-order Fourier series have been neglected (valid if $\pi^2 \tilde{\delta} \gg 1$). This solution is actually unphysical, since the energy peaks occur when DW is at cell edges if $\tilde{\delta}\gg 1$ (but at cell centers if $\tilde{\delta} \ll 1$). To correct this mismatch, we derive another discretized form of total energy $F = \sum_n \frac{1}{2}b (P_0^2 - P_n^2)^2$ \footnote{The key step of this derivation is to use $P_n - P_{n-1} \approx P_0(1-(\frac{P_n}{P_0})^2) \frac{h_l}{\delta}$ (using the Taylor expansion of the tanh function) to calculate the gradient energy $\frac{\kappa}{2 h_l^2} \sum_n (P_{n+1} - P_n)^2$}, and then arrive at a new expression of   $\gamma(x_{DW})$:
\begin{equation}
    \gamma_{180,ana2} = \gamma_0 [1 + 2\pi^2 \tilde{\delta} (1 + \pi^2 \tilde{\delta}^2) \mathrm{csch}(\pi^2\tilde{\delta}) \cos (2\pi \tilde{x}_{DW})   ],
    \label{eq:gamma_mine}
\end{equation}
which gives correct locations of extremums. Similarly, for ``90$\searrow \nearrow$" DW with $\eta = 3b$, we get $\gamma_{90,\searrow \nearrow, ana} = \frac{1}{2} \gamma_{180,ana}$ for both analytical solutions.

The comparison between $\gamma_{180,ana2}$,  $\gamma_{90,\searrow \nearrow, ana2}$ and the numerical solution is shown in Figure \ref{fig:figure1}(b). We can see that our analytical solution gives qualitatively correct pictures, though error still exists (even if we add more Fourier series) since the analytical solution has used the continuum solution of $P$ in the derivation. 

Once we get $\gamma(x_{DW})$, we can estimate the TEF ($E_{th}$) using $E_{th,ana} = \frac{1}{\cos\theta \Delta P} |\pd{\gamma}{x_{DW}} |_{max} $ ($\Delta P = 2P_0, \sqrt{2} P_0, \sqrt{2}P_0$ and $\theta = 0, 0, \pi/4$ for 180,  ``90$\searrow \nearrow$", ``90$\rightarrow \uparrow$" DW, respectively). The results indicate that 
$E_{th,ana}$ decays to zero as $\tilde{\delta}$ goes to infinity, and $E_{th, 90\searrow \nearrow, ana} = \frac{1}{\sqrt{2}}E_{th,180,ana}$ if $\eta = 3b$, which trends are consistent with the numerical solution of $E_{th}$, as shown in Figure \ref{fig:figure2}(a). However, Equation \ref{eq:gamma_cahn} underestimates TEF while Equation \ref{eq:gamma_mine} overestimates TEF. In addition, the numerical solution also shows that $E_{th}$ for DW motion approaches that for uniform switching as $\delta$ goes to zero, which is not captured by Equation \ref{eq:gamma_cahn} or Equation \ref{eq:gamma_mine}. Therefore, we propose the following empirical model for the entire range of $\tilde{\delta}$:
\begin{equation}
    E_{th,emp} = E_{th,uni} e^{-\alpha \tilde{\delta}^2},
    \label{eq:Eth_emp}
\end{equation}
where $E_{th,uni} = E_0, \frac{1}{\sqrt{2}} E_0, E_0$, and $\alpha = \sqrt{2\pi}, \sqrt{2\pi}, 2\sqrt{2\pi}$ for 180, ``90 $\searrow \nearrow$", ``90 $\rightarrow \uparrow$" DWs \rv{shown in Figure \ref{fig:figure2}(a)}, respectively. This model fits numerical results quite well.  ``90 $\rightarrow \uparrow$" DW has smaller TEF than ``90 $\searrow \nearrow$", due to the  distortion of electric field. Meanwhile, Figure \ref{fig:figure2}(b) shows that the stable DW energy ($\gamma^* = \min(\gamma)$) monotonically increases with $\tilde{\delta}$, and is overestimated by the continuum theory at small $\tilde{\delta}$.  In addition, as shown in Figure \ref{fig:figure2}(c), the mobility ($\mu \doteq v_{DW,x}/E_{app}$) of 180 and ``90$\searrow \nearrow$" DW goes to zero for small $E_{app}$ and small $\tilde{\delta}$, and approaches the continuum theory for large $E$ and large $\delta$.  ``90 $\rightarrow \uparrow$" DW moves faster than that predicted by continuum theory even for thick DWs.  

The trend of increasing $E_{th}$ with decreasing $\gamma^*$ and $\delta$ shown in  Figure \ref{fig:figure2} for 180 DW and ``90 $\rightarrow \uparrow$" DW is roughly consistent with the existing first-principles studies. For HfO$_2$, it has been shown that the  switching barrier of near-zero-energy flat-phonon-band 180$^\circ$ DW is close to that of uniform switching \cite{lee_scale-free_2020}, and that of the high-energy quasi-chiral 180$^\circ$ DW ($\tilde{\delta}\in [0.5, 1]$) is small and comparable with experimental observation ($E_{th}/E_0\sim 0.15$, stress-free condition) \cite{choe_unexpectedly_2021} . For PbTiO$_3$, it has been show that 90$^\circ$ DW has a much smaller energy \cite{meyer_ab_2002} and TEF \cite{liu_intrinsic_2016} compared with 180$^\circ$ DWs. This indicates that a two-step sequential 90$^\circ$ DW motion may be energetically more favorable compared with one 180$^\circ$ DW motion. Note that  the approach presented in this \rv{section} can be extended to \rv{more complex} $f_{landau}$ to better fit the first-principles results.

\begin{figure}[t!]
    \centering
    \includegraphics[width=8.6cm]{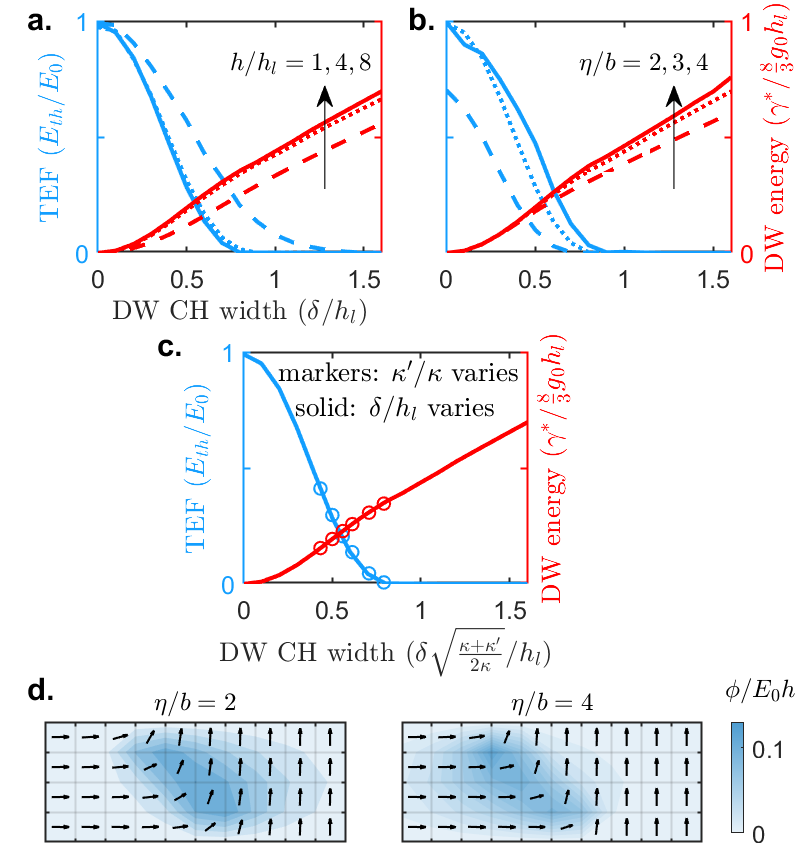}
    \caption{The effect of (a) film thickness, (b)  anisotropic \rv{permittivity} factor $\eta/b$\rv{, and (c) anisotropic gradient factor $\kappa'/\kappa$} on the threshold electric field (TEF) and energy of ``90$\rightarrow \uparrow$" DW, and \rv{(d)} the effect of $\eta/b$ on local polarization and electric potential. \rv{In (c),  $\kappa'=\kappa$ as $\delta/h_l$ varies, and $\delta/h_l = 0.5$ as  $\kappa'/\kappa$ varies.} }
    \label{fig:figure3}
\end{figure}

\rv{\subsection{Supplementary results for DW motion}}

Next, we present the effect of film thickness $h$, anisotropic \rv{permittivity} factor $\eta/3b$, \rv{anisotropic gradient factor $\kappa'/\kappa$, }and background permittivity $\epsilon_{r,b}$ for ``90 $\rightarrow \uparrow$" DW. These parameters do not influence the 180$^\circ$ DW, and $h$\rv{, $\epsilon_{r,b}$ do} not influence the ``90 $\searrow \nearrow$" DW. \rv{Unless stated, all the parameters are the same as the base case shown in Figure \ref{fig:figure2}. }As already shown  in Figure \ref{fig:figure1}(a), the electric field is distorted near electrodes. Figure \ref{fig:figure3}(a) shows that this distortion does not change TEF and DW energy much for films thicker than 4$h_l$ (typically $\sim$2 nm).   Figure \ref{fig:figure3}(b) shows that smaller $\eta/b$ leads to smaller TEF and $\gamma^*$, which can be explained by the asymmetric electric field on the two sides of DWs as shown in Figure \ref{fig:figure3}\rv{(d)}. Note that the above trend can be reversed for $E_{app}$ in the other direction. \rv{The including of anisotropic gradient factor $\frac{\kappa'}{\kappa}$ does not change the relation between TEF and DW width, as long as we use the correct  90$^\circ$ DW half width $\delta \sqrt{\frac{\kappa + \kappa'}{2\kappa}}$ when $\kappa\neq \kappa'$, as shown in Figure \ref{fig:figure3}(c).} The background permittivity has negligible effect on TEF and DW energy (not shown in Figure \ref{fig:figure3}).

\rv{\section{Model and results of a real ferroelectric thin film}}
\rv{In Sec.II, we consider general ferroelectric second-order phase transition without elastic energy. The simple model makes it easy to draw revealing conclusions.  However, for practical use, it may be necessary to use more complex models which are well-parameterized for real ferroelectric materials. In this section, we discuss on the role of elastic energy, simulate a real ferroelectric material, BaTiO$_3$, and check the validity of the conclusions drawn in Sec. II.}

\rv{First, we briefly review the effect of elastic energy on DWs in ferroelectric bulk materials and thin films. (1) For infinitely large isolated DW in bulk materials with stress-free conditions in the homogeneous region ($\sigma_{xy} = 0$ at any location; $\sigma_{xx} =  \sigma_{yy}=0$, $P_y = \pm P_0$, $P_x = 0$ for 180$^\circ$ DW or $P_y = \pm P_0/\sqrt{2}$, $P_x = P_0/\sqrt{2}$ for 90$^\circ$ DW far away from DWs), the effect of elastic energy can be included by simply modifying Landau parameters ($a$ and $b$). This point has been thoroughly discussed in previous papers \cite{cao_theory_1991, hlinka_phenomenological_2006}. In this case, the including of elastic energy does not change the conclusions drawn in Sec. II as long as we use the corrected Landau parameters. (2) In ferroelectric thin films, the misfit strain can influence the equilibrium domain structures \cite{li_effect_2002}, and there may not exist analytical solutions. In the following, we directly simulate a BaTiO$_3$ thin film to study the DW dynamics. In the simulations, we fix the initial locations of DWs. If the TEF is low enough, the system can go to its equilibrium state where the DWs may even disappear. Since we only care about large enough TEF, this setting should still be reasonable. But we should be aware that the initial condition and simulation area can influence the results.   }

\rv{For the simulations, we use the normalized phase-field model of BaTiO$_3$ given by \cite{zhang_computational_2005}, where all the variables are dimensionless (the corresponding dimensional variables can be calculated accordingly). Compared with the model in Sec.II, this model uses more polynomial terms ($P_x^6$, $P_y^6$, $P_x^4 P_y^4$) in $f_{landau}$, includes elastic energy, and assumes isotropic gradient tenor. Note that as pointed out by \cite{hlinka_phenomenological_2006}, BaTiO$_3$ should have an anisotropic gradient tensor ($\kappa'/\kappa = 51/2$). The computation area is $180h_l \times 40 h_l$. In terms of the boundary conditions, we assume prescribed displacement at the bottom, stress free condition at the top, and periodic boundary condition at the two sides \cite{li_effect_2002}. In the initial state, we put two complementary DWs at the one-quarter and three-quarter points in $x$ direction, to satisfy the periodic boundary condition. We define the misfit strain as $\Delta \varepsilon_S = \varepsilon_S - \varepsilon_0$, where $\varepsilon_0 = -0.44\%$ for 180$^\circ$ DW, and $\varepsilon_0 = 0.105\%$ for 90$\nearrow \searrow$ DW. Here we do not consider $90\rightarrow \uparrow$ DW, to keep the strain as a constant far from DWs. We still use finite volume method with rectangular mesh for discretization, and we change mesh size $h_l = 5, 10, 15, 20$ to check if Equation \ref{eq:Eth_emp} still works. Note that the mesh size should be close to real lattice size if we set $h_l\approx 8.2$, since the model in \cite{zhang_computational_2005} gives 180 DW width as $\approx 13.6$ (using Equation 4.10 in \cite{cao_theory_1991}), while in dimensional units BaTiO$_3$ has unit cell size $\approx$ 0.4 nm and 180 DW width $\approx$ 0.67 nm \cite{hlinka_phenomenological_2006}.   }

\begin{figure}
    \centering
    \includegraphics[width=0.95\linewidth]{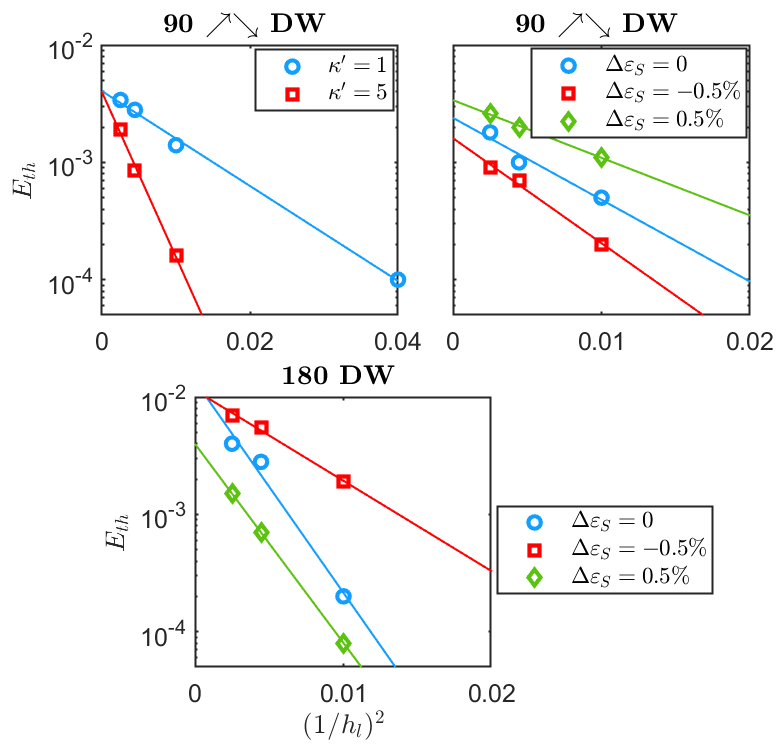}
    \caption{\rv{The threshold electrc field ($E_{th}$) to drive (1) 90$^\circ$ DWs with different gradient coefficient $\kappa'$, (2) 90$^\circ$ DWs with different misfit strain, (3) 180$^\circ$ DWs with different misfit strain.} }
    \label{fig:figure4}
\end{figure}

\rv{The simulation results are summarized in Figure \ref{fig:figure4}. First, we can see that all the results roughly follow Equation \ref{eq:Eth_emp} that $\ln(E_{th})\sim -(1/h_l)^2$. This proves that for such a complex model with elastic energy, the conclusion drawn in Sec. II that the TEF increases with decreasing $\delta/h_l$ still holds true. Next, we notice that the 180$^\circ$ DW has smaller TEF than the 90$^\circ$ DW when $\Delta \varepsilon_S = 0$ for the model given by \cite{zhang_computational_2005} ($\kappa' = \kappa = 1$). However, actually BaTiO$_3$ should have a very large  $\kappa'/\kappa$, leading to a very thick 90$^\circ$ DW (over 3 nm \cite{hlinka_phenomenological_2006}) which should have negligible $E_{th}$. This can be supported by Figure \ref{fig:figure4}(a), since $\kappa'=5$ has already made $E_{th}< 1\% E_0$ at $h_l = 8.2$. This is a good example that the anisotropic gradient tensor can be very important to calculate TEF for 90$^\circ$ DW. Finally, Figure \ref{fig:figure4}(b)(c) proves that the misfit strain can influence the TEF of DWs. A small compressive misfit strain increases TEF of 180$^\circ$ DWs since it increases the uniform switching barrier \cite{zhang_computational_2005}, but decreases TEF of 90$^\circ$ DWs possibly because it destabilizes 90$^\circ$ DWs \cite{li_effect_2002}.     }

\rv{\section{Discussion}}
\rv{In Sec.II and Sec.III, we have used discretized phase-field models to show that the intrinsic TEF to drive DWs in ferroelectric materials increases as the ratio of DW width to unit cell size decreases, and the TEF becomes significant when the DW is thinner than two unit cells. The continuum model gives zero TEF, which is a good approximation only for very thick DWs. Then we can make two inferences. First, Equation \ref{eq:Eth_emp} can be used to roughly judge whether TEF of a DW in a specific ferroelectric material is significant. For example, for BaTiO$_3$, the 90$^\circ$ DW width is over 3 nm ($\sim2\delta$) at room temperature, while the unit cell size is only around 0.4 nm, resulting in a very small TEF ($<10^{-4} E_0$) which can be safely ignored in most cases. Meanwhile, the 180$^\circ$ DW in BaTiO$_3$ has thickness around 0.67 nm (room temperature), leading to TEF around $0.1 E_0$, which cannot be neglected if the applied electric field is on the order of $0.1 E_0$. In addition, some 180$^\circ$ DWs in HfO$_2$ have thickness smaller than two unit cells, which should also have significant TEF (which has been proven in first-principles calculations \cite{choe_unexpectedly_2021, lee_scale-free_2020}).  The second inference is, Equation \ref{eq:Eth_emp} can also be used to judge whether the mesh size is reasonable in numerical simulations. Mesh elements much larger (smaller) than unit cells can overestimate (underestimate) the TEF of DW, which may give misleading simulation results, especially when the applied electric field is small. }

\begin{figure}
    \centering
    \includegraphics[width=0.95\linewidth]{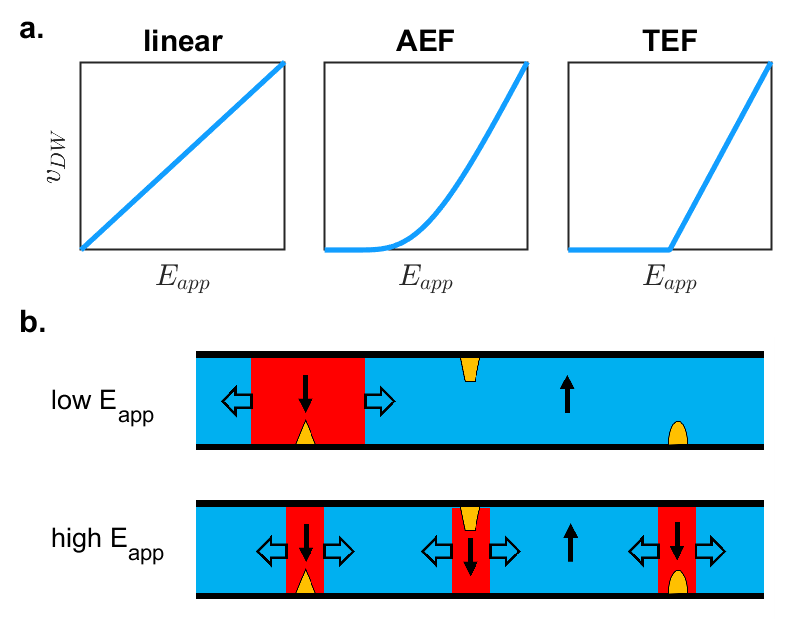}
    \caption{\rv{(a) Three possible relations between (effective) the DW velocity and the applied electic field. (b) A possible explanation of the AEF behavior. The solid arrows represent the polarization direction, and the hollow arrows represent the direction of DW motion. } }
    \label{fig:figure5}
\end{figure}

\rv{Then we discuss the temperature effect. For the model in Sec.II (second-order phase transition without elastic energy), it is usually assumed that $a = a_0(T_c-T)$, where $T_c$ is the Curie temperature; and other parameters $b, \eta, \kappa, \kappa', \tau$ are independent of temperature \cite{chandra_landau_2007}. Then we can derive  $\delta \sim |T_c - T|^{-1/2}$, $E_0 \sim |T_c-T|^{3/2}$, and Equation \ref{eq:Eth_emp} can be re-written as $E_{th,emp} \sim |T_c-T|^{3/2} \exp(-\alpha \tilde{\delta}_{0K}^2 \frac{T_c}{T_c - T} )$. Therefore, the TEF of a DW reduces quickly as the temperature increases. More complex phase-field models should give similar trends. Therefore, the DW velocity versus applied electric field should show nearly linear relation at high temperature and TEF behavior at low temperature (as shown in Figure \ref{fig:figure5}(a)). This is also consistent with first-principle studies \cite{liu_intrinsic_2016}. }

\rv{Next, we try to relate the DW velocity to the switching speed of a whole ferroelectric thin film (switching speed in short). On the one hand, both the velocity of an isolated DW and the switching speed show TEF behaviors at low temperature. It is very possible that the former is the main cause for the latter.  On the other hand, at high temperature, the velocity of an isolated DW has a linear dependence on the applied electric field, while the switching speed has a nonlinear, AEF-like dependence (the difference is shown in Figure \ref{fig:figure5}(a)). Therefore, the single DW model cannot explain the AEF of the switching speed.} 

\rv{Here we provide one possible explanation of the AEF behavior, as shown in Figure \ref{fig:figure5}(b). We propose that the intrinsic defects (such as grain boundaries, phase boundaries, oxygen vacancies, surface roughness) in real ferroelectric films give a distribution of TEF to nucleate reverse domains. At high temperature, the TEF of DW motion is smaller than the TEF of domain nucleation. In this case, there should be more DWs (or reverse domains) at the early stage of switching at higher $E_{app}$. Since the switching speed is roughly proportional to the number of DWs times the DW velocity, the switching speed turns out to have a nonlinear dependence on $E_{app}$ and shows AEF-like behaviors. Furthermore, the TEF of DW motion may be larger than the/some TEFs of domain nucleation at low temperature. In this case, the number of DWs may not change much with $E_{app}$  as long as $E_{app}>E_{th}$, and the switching speed shows TEF-like behaviors. This conceptual model unifies the view of AEF and TEF, and gives a new explanation of AEF beyond thermal nucleation theory. A rigorous mathematical model with careful and reasonable treatment of the defects need to be developed to prove this conceptual model in future works.  }

\rv{\section{conclusion}}

\rv{In summary, we investigate the threshold electric field (TEF) for a wide range of DW width and energy for 180$^\circ$ and 90$^\circ$ DWs, based on discretized phase-field models. We present numerical results, analytical approximations, and empirical model for the TEF of flat DWs, which is qualitatively consistent with existing first-principles studies.  Furthermore, our results demonstrate that the mesh size and mesh orientation in phase-field simulations must be carefully treated when studying the low-field switching behaviors. We also propose a conceptual model for AEF which need to be rigorously proved in future works. }

\begin{acknowledgments}
This work was supported in part by the NSFC under Grants  92164204, 62222119.
\end{acknowledgments}


\bibliography{apssamp}

\begin{thebibliography}{38}%
\makeatletter
\providecommand \@ifxundefined [1]{%
 \@ifx{#1\undefined}
}%
\providecommand \@ifnum [1]{%
 \ifnum #1\expandafter \@firstoftwo
 \else \expandafter \@secondoftwo
 \fi
}%
\providecommand \@ifx [1]{%
 \ifx #1\expandafter \@firstoftwo
 \else \expandafter \@secondoftwo
 \fi
}%
\providecommand \natexlab [1]{#1}%
\providecommand \enquote  [1]{``#1''}%
\providecommand \bibnamefont  [1]{#1}%
\providecommand \bibfnamefont [1]{#1}%
\providecommand \citenamefont [1]{#1}%
\providecommand \href@noop [0]{\@secondoftwo}%
\providecommand \href [0]{\begingroup \@sanitize@url \@href}%
\providecommand \@href[1]{\@@startlink{#1}\@@href}%
\providecommand \@@href[1]{\endgroup#1\@@endlink}%
\providecommand \@sanitize@url [0]{\catcode `\\12\catcode `\$12\catcode `\&12\catcode `\#12\catcode `\^12\catcode `\_12\catcode `\%12\relax}%
\providecommand \@@startlink[1]{}%
\providecommand \@@endlink[0]{}%
\providecommand \url  [0]{\begingroup\@sanitize@url \@url }%
\providecommand \@url [1]{\endgroup\@href {#1}{\urlprefix }}%
\providecommand \urlprefix  [0]{URL }%
\providecommand \Eprint [0]{\href }%
\providecommand \doibase [0]{https://doi.org/}%
\providecommand \selectlanguage [0]{\@gobble}%
\providecommand \bibinfo  [0]{\@secondoftwo}%
\providecommand \bibfield  [0]{\@secondoftwo}%
\providecommand \translation [1]{[#1]}%
\providecommand \BibitemOpen [0]{}%
\providecommand \bibitemStop [0]{}%
\providecommand \bibitemNoStop [0]{.\EOS\space}%
\providecommand \EOS [0]{\spacefactor3000\relax}%
\providecommand \BibitemShut  [1]{\csname bibitem#1\endcsname}%
\let\auto@bib@innerbib\@empty
\bibitem [{\citenamefont {Schroeder}\ \emph {et~al.}(2022)\citenamefont {Schroeder}, \citenamefont {Park}, \citenamefont {Mikolajick},\ and\ \citenamefont {Hwang}}]{schroeder_fundamentals_2022}%
  \BibitemOpen
  \bibfield  {author} {\bibinfo {author} {\bibfnamefont {U.}~\bibnamefont {Schroeder}}, \bibinfo {author} {\bibfnamefont {M.~H.}\ \bibnamefont {Park}}, \bibinfo {author} {\bibfnamefont {T.}~\bibnamefont {Mikolajick}},\ and\ \bibinfo {author} {\bibfnamefont {C.~S.}\ \bibnamefont {Hwang}},\ }\bibfield  {title} {\bibinfo {title} {The fundamentals and applications of ferroelectric {HfO2}},\ }\href {https://doi.org/10.1038/s41578-022-00431-2} {\bibfield  {journal} {\bibinfo  {journal} {Nat Rev Mater}\ }\textbf {\bibinfo {volume} {7}},\ \bibinfo {pages} {653} (\bibinfo {year} {2022})}\BibitemShut {NoStop}%
\bibitem [{\citenamefont {Ramaswamy}\ \emph {et~al.}(2023)\citenamefont {Ramaswamy}, \citenamefont {Calderoni}, \citenamefont {Zahurak}, \citenamefont {Servalli}, \citenamefont {Chavan}, \citenamefont {Chhajed}, \citenamefont {Balakrishnan}, \citenamefont {Fischer}, \citenamefont {Hollander}, \citenamefont {Ettisserry}, \citenamefont {Liao}, \citenamefont {Karda}, \citenamefont {Jerry}, \citenamefont {Mariani}, \citenamefont {Visconti}, \citenamefont {Cook}, \citenamefont {Cook}, \citenamefont {Mills}, \citenamefont {Torsi}, \citenamefont {Mouli}, \citenamefont {Byers}, \citenamefont {Helm}, \citenamefont {Pawlowski}, \citenamefont {Shiratake},\ and\ \citenamefont {Chandrasekaran}}]{ramaswamy_nvdram_2023}%
  \BibitemOpen
  \bibfield  {author} {\bibinfo {author} {\bibfnamefont {N.}~\bibnamefont {Ramaswamy}}, \bibinfo {author} {\bibfnamefont {A.}~\bibnamefont {Calderoni}}, \bibinfo {author} {\bibfnamefont {J.}~\bibnamefont {Zahurak}}, \bibinfo {author} {\bibfnamefont {G.}~\bibnamefont {Servalli}}, \bibinfo {author} {\bibfnamefont {A.}~\bibnamefont {Chavan}}, \bibinfo {author} {\bibfnamefont {S.}~\bibnamefont {Chhajed}}, \bibinfo {author} {\bibfnamefont {M.}~\bibnamefont {Balakrishnan}}, \bibinfo {author} {\bibfnamefont {M.}~\bibnamefont {Fischer}}, \bibinfo {author} {\bibfnamefont {M.}~\bibnamefont {Hollander}}, \bibinfo {author} {\bibfnamefont {D.~P.}\ \bibnamefont {Ettisserry}}, \bibinfo {author} {\bibfnamefont {A.}~\bibnamefont {Liao}}, \bibinfo {author} {\bibfnamefont {K.}~\bibnamefont {Karda}}, \bibinfo {author} {\bibfnamefont {M.}~\bibnamefont {Jerry}}, \bibinfo {author} {\bibfnamefont {M.}~\bibnamefont {Mariani}}, \bibinfo {author} {\bibfnamefont {A.}~\bibnamefont {Visconti}}, \bibinfo {author} {\bibfnamefont {B.~R.}\
  \bibnamefont {Cook}}, \bibinfo {author} {\bibfnamefont {B.~D.}\ \bibnamefont {Cook}}, \bibinfo {author} {\bibfnamefont {D.}~\bibnamefont {Mills}}, \bibinfo {author} {\bibfnamefont {A.}~\bibnamefont {Torsi}}, \bibinfo {author} {\bibfnamefont {C.}~\bibnamefont {Mouli}}, \bibinfo {author} {\bibfnamefont {E.}~\bibnamefont {Byers}}, \bibinfo {author} {\bibfnamefont {M.}~\bibnamefont {Helm}}, \bibinfo {author} {\bibfnamefont {S.}~\bibnamefont {Pawlowski}}, \bibinfo {author} {\bibfnamefont {S.}~\bibnamefont {Shiratake}},\ and\ \bibinfo {author} {\bibfnamefont {N.}~\bibnamefont {Chandrasekaran}},\ }\bibfield  {title} {\bibinfo {title} {{NVDRAM}: {A} {32Gb} {Dual} {Layer} {3D} {Stacked} {Non}-volatile {Ferroelectric} {Memory} with {Near}-{DRAM} {Performance} for {Demanding} {AI} {Workloads}},\ }in\ \href {https://doi.org/10.1109/IEDM45741.2023.10413848} {\emph {\bibinfo {booktitle} {2023 {International} {Electron} {Devices} {Meeting} ({IEDM})}}}\ (\bibinfo  {publisher} {IEEE},\ \bibinfo {address} {San Francisco, CA,
  USA},\ \bibinfo {year} {2023})\ pp.\ \bibinfo {pages} {1--4}\BibitemShut {NoStop}%
\bibitem [{\citenamefont {Silva}\ \emph {et~al.}(2023)\citenamefont {Silva}, \citenamefont {Alcala}, \citenamefont {Avci}, \citenamefont {Barrett}, \citenamefont {Bégon-Lours}, \citenamefont {Borg}, \citenamefont {Byun}, \citenamefont {Chang}, \citenamefont {Cheong}, \citenamefont {Choe}, \citenamefont {Coignus}, \citenamefont {Deshpande}, \citenamefont {Dimoulas}, \citenamefont {Dubourdieu}, \citenamefont {Fina}, \citenamefont {Funakubo}, \citenamefont {Grenouillet}, \citenamefont {Gruverman}, \citenamefont {Heo}, \citenamefont {Hoffmann}, \citenamefont {Hsain}, \citenamefont {Huang}, \citenamefont {Hwang}, \citenamefont {Íñiguez}, \citenamefont {Jones}, \citenamefont {Karpov}, \citenamefont {Kersch}, \citenamefont {Kwon}, \citenamefont {Lancaster}, \citenamefont {Lederer}, \citenamefont {Lee}, \citenamefont {Lomenzo}, \citenamefont {Martin}, \citenamefont {Martin}, \citenamefont {Migita}, \citenamefont {Mikolajick}, \citenamefont {Noheda}, \citenamefont {Park}, \citenamefont {Rabe}, \citenamefont
  {Salahuddin}, \citenamefont {Sánchez}, \citenamefont {Seidel}, \citenamefont {Shimizu}, \citenamefont {Shiraishi}, \citenamefont {Slesazeck}, \citenamefont {Toriumi}, \citenamefont {Uchida}, \citenamefont {Vilquin}, \citenamefont {Xu}, \citenamefont {Ye},\ and\ \citenamefont {Schroeder}}]{silva_roadmap_2023}%
  \BibitemOpen
  \bibfield  {author} {\bibinfo {author} {\bibfnamefont {J.~P.~B.}\ \bibnamefont {Silva}}, \bibinfo {author} {\bibfnamefont {R.}~\bibnamefont {Alcala}}, \bibinfo {author} {\bibfnamefont {U.~E.}\ \bibnamefont {Avci}}, \bibinfo {author} {\bibfnamefont {N.}~\bibnamefont {Barrett}}, \bibinfo {author} {\bibfnamefont {L.}~\bibnamefont {Bégon-Lours}}, \bibinfo {author} {\bibfnamefont {M.}~\bibnamefont {Borg}}, \bibinfo {author} {\bibfnamefont {S.}~\bibnamefont {Byun}}, \bibinfo {author} {\bibfnamefont {S.-C.}\ \bibnamefont {Chang}}, \bibinfo {author} {\bibfnamefont {S.-W.}\ \bibnamefont {Cheong}}, \bibinfo {author} {\bibfnamefont {D.-H.}\ \bibnamefont {Choe}}, \bibinfo {author} {\bibfnamefont {J.}~\bibnamefont {Coignus}}, \bibinfo {author} {\bibfnamefont {V.}~\bibnamefont {Deshpande}}, \bibinfo {author} {\bibfnamefont {A.}~\bibnamefont {Dimoulas}}, \bibinfo {author} {\bibfnamefont {C.}~\bibnamefont {Dubourdieu}}, \bibinfo {author} {\bibfnamefont {I.}~\bibnamefont {Fina}}, \bibinfo {author} {\bibfnamefont
  {H.}~\bibnamefont {Funakubo}}, \bibinfo {author} {\bibfnamefont {L.}~\bibnamefont {Grenouillet}}, \bibinfo {author} {\bibfnamefont {A.}~\bibnamefont {Gruverman}}, \bibinfo {author} {\bibfnamefont {J.}~\bibnamefont {Heo}}, \bibinfo {author} {\bibfnamefont {M.}~\bibnamefont {Hoffmann}}, \bibinfo {author} {\bibfnamefont {H.~A.}\ \bibnamefont {Hsain}}, \bibinfo {author} {\bibfnamefont {F.-T.}\ \bibnamefont {Huang}}, \bibinfo {author} {\bibfnamefont {C.~S.}\ \bibnamefont {Hwang}}, \bibinfo {author} {\bibfnamefont {J.}~\bibnamefont {Íñiguez}}, \bibinfo {author} {\bibfnamefont {J.~L.}\ \bibnamefont {Jones}}, \bibinfo {author} {\bibfnamefont {I.~V.}\ \bibnamefont {Karpov}}, \bibinfo {author} {\bibfnamefont {A.}~\bibnamefont {Kersch}}, \bibinfo {author} {\bibfnamefont {T.}~\bibnamefont {Kwon}}, \bibinfo {author} {\bibfnamefont {S.}~\bibnamefont {Lancaster}}, \bibinfo {author} {\bibfnamefont {M.}~\bibnamefont {Lederer}}, \bibinfo {author} {\bibfnamefont {Y.}~\bibnamefont {Lee}}, \bibinfo {author} {\bibfnamefont
  {P.~D.}\ \bibnamefont {Lomenzo}}, \bibinfo {author} {\bibfnamefont {L.~W.}\ \bibnamefont {Martin}}, \bibinfo {author} {\bibfnamefont {S.}~\bibnamefont {Martin}}, \bibinfo {author} {\bibfnamefont {S.}~\bibnamefont {Migita}}, \bibinfo {author} {\bibfnamefont {T.}~\bibnamefont {Mikolajick}}, \bibinfo {author} {\bibfnamefont {B.}~\bibnamefont {Noheda}}, \bibinfo {author} {\bibfnamefont {M.~H.}\ \bibnamefont {Park}}, \bibinfo {author} {\bibfnamefont {K.~M.}\ \bibnamefont {Rabe}}, \bibinfo {author} {\bibfnamefont {S.}~\bibnamefont {Salahuddin}}, \bibinfo {author} {\bibfnamefont {F.}~\bibnamefont {Sánchez}}, \bibinfo {author} {\bibfnamefont {K.}~\bibnamefont {Seidel}}, \bibinfo {author} {\bibfnamefont {T.}~\bibnamefont {Shimizu}}, \bibinfo {author} {\bibfnamefont {T.}~\bibnamefont {Shiraishi}}, \bibinfo {author} {\bibfnamefont {S.}~\bibnamefont {Slesazeck}}, \bibinfo {author} {\bibfnamefont {A.}~\bibnamefont {Toriumi}}, \bibinfo {author} {\bibfnamefont {H.}~\bibnamefont {Uchida}}, \bibinfo {author} {\bibfnamefont
  {B.}~\bibnamefont {Vilquin}}, \bibinfo {author} {\bibfnamefont {X.}~\bibnamefont {Xu}}, \bibinfo {author} {\bibfnamefont {K.~H.}\ \bibnamefont {Ye}},\ and\ \bibinfo {author} {\bibfnamefont {U.}~\bibnamefont {Schroeder}},\ }\bibfield  {title} {\bibinfo {title} {Roadmap on ferroelectric hafnia- and zirconia-based materials and devices},\ }\href {https://doi.org/10.1063/5.0148068} {\bibfield  {journal} {\bibinfo  {journal} {APL Materials}\ }\textbf {\bibinfo {volume} {11}},\ \bibinfo {pages} {089201} (\bibinfo {year} {2023})}\BibitemShut {NoStop}%
\bibitem [{\citenamefont {Böscke}\ \emph {et~al.}(2011)\citenamefont {Böscke}, \citenamefont {Müller}, \citenamefont {Bräuhaus}, \citenamefont {Schröder},\ and\ \citenamefont {Böttger}}]{boscke_ferroelectricity_2011}%
  \BibitemOpen
  \bibfield  {author} {\bibinfo {author} {\bibfnamefont {T.~S.}\ \bibnamefont {Böscke}}, \bibinfo {author} {\bibfnamefont {J.}~\bibnamefont {Müller}}, \bibinfo {author} {\bibfnamefont {D.}~\bibnamefont {Bräuhaus}}, \bibinfo {author} {\bibfnamefont {U.}~\bibnamefont {Schröder}},\ and\ \bibinfo {author} {\bibfnamefont {U.}~\bibnamefont {Böttger}},\ }\bibfield  {title} {\bibinfo {title} {Ferroelectricity in hafnium oxide thin films},\ }\href {https://doi.org/10.1063/1.3634052} {\bibfield  {journal} {\bibinfo  {journal} {Applied Physics Letters}\ }\textbf {\bibinfo {volume} {99}},\ \bibinfo {pages} {102903} (\bibinfo {year} {2011})}\BibitemShut {NoStop}%
\bibitem [{\citenamefont {Wang}\ \emph {et~al.}(2023)\citenamefont {Wang}, \citenamefont {Tao}, \citenamefont {Guzman}, \citenamefont {Luo}, \citenamefont {Zhou}, \citenamefont {Yang}, \citenamefont {Wei}, \citenamefont {Liu}, \citenamefont {Jiang}, \citenamefont {Chen}, \citenamefont {Lv}, \citenamefont {Ding}, \citenamefont {Wei}, \citenamefont {Gong}, \citenamefont {Wang}, \citenamefont {Liu}, \citenamefont {Du},\ and\ \citenamefont {Liu}}]{wang_stable_2023}%
  \BibitemOpen
  \bibfield  {author} {\bibinfo {author} {\bibfnamefont {Y.}~\bibnamefont {Wang}}, \bibinfo {author} {\bibfnamefont {L.}~\bibnamefont {Tao}}, \bibinfo {author} {\bibfnamefont {R.}~\bibnamefont {Guzman}}, \bibinfo {author} {\bibfnamefont {Q.}~\bibnamefont {Luo}}, \bibinfo {author} {\bibfnamefont {W.}~\bibnamefont {Zhou}}, \bibinfo {author} {\bibfnamefont {Y.}~\bibnamefont {Yang}}, \bibinfo {author} {\bibfnamefont {Y.}~\bibnamefont {Wei}}, \bibinfo {author} {\bibfnamefont {Y.}~\bibnamefont {Liu}}, \bibinfo {author} {\bibfnamefont {P.}~\bibnamefont {Jiang}}, \bibinfo {author} {\bibfnamefont {Y.}~\bibnamefont {Chen}}, \bibinfo {author} {\bibfnamefont {S.}~\bibnamefont {Lv}}, \bibinfo {author} {\bibfnamefont {Y.}~\bibnamefont {Ding}}, \bibinfo {author} {\bibfnamefont {W.}~\bibnamefont {Wei}}, \bibinfo {author} {\bibfnamefont {T.}~\bibnamefont {Gong}}, \bibinfo {author} {\bibfnamefont {Y.}~\bibnamefont {Wang}}, \bibinfo {author} {\bibfnamefont {Q.}~\bibnamefont {Liu}}, \bibinfo {author} {\bibfnamefont
  {S.}~\bibnamefont {Du}},\ and\ \bibinfo {author} {\bibfnamefont {M.}~\bibnamefont {Liu}},\ }\bibfield  {title} {\bibinfo {title} {A stable rhombohedral phase in ferroelectric {Hf}({Zr})1+{xO2} capacitor with ultralow coercive field},\ }\href@noop {} {\bibfield  {journal} {\bibinfo  {journal} {Science}\ }\textbf {\bibinfo {volume} {381}},\ \bibinfo {pages} {558} (\bibinfo {year} {2023})}\BibitemShut {NoStop}%
\bibitem [{\citenamefont {Deng}\ \emph {et~al.}(2023)\citenamefont {Deng}, \citenamefont {Xiao}, \citenamefont {Zhao}, \citenamefont {Huang}, \citenamefont {Kampfe}, \citenamefont {Narayanan},\ and\ \citenamefont {Ni}}]{deng_comparative_2023}%
  \BibitemOpen
  \bibfield  {author} {\bibinfo {author} {\bibfnamefont {S.}~\bibnamefont {Deng}}, \bibinfo {author} {\bibfnamefont {Y.}~\bibnamefont {Xiao}}, \bibinfo {author} {\bibfnamefont {Z.}~\bibnamefont {Zhao}}, \bibinfo {author} {\bibfnamefont {T.-J.}\ \bibnamefont {Huang}}, \bibinfo {author} {\bibfnamefont {T.}~\bibnamefont {Kampfe}}, \bibinfo {author} {\bibfnamefont {V.}~\bibnamefont {Narayanan}},\ and\ \bibinfo {author} {\bibfnamefont {K.}~\bibnamefont {Ni}},\ }\bibfield  {title} {\bibinfo {title} {Comparative {Advantages} of {2T}-{nC} {FeRAM} in {Empowering} {High} {Density} {3D} {Ferroelectric} {Capacitor} {Memory}},\ }in\ \href {https://doi.org/10.1109/IEDM45741.2023.10413855} {\emph {\bibinfo {booktitle} {2023 {International} {Electron} {Devices} {Meeting} ({IEDM})}}}\ (\bibinfo  {publisher} {IEEE},\ \bibinfo {address} {San Francisco, CA, USA},\ \bibinfo {year} {2023})\ pp.\ \bibinfo {pages} {1--4}\BibitemShut {NoStop}%
\bibitem [{\citenamefont {Fu}\ \emph {et~al.}(2023)\citenamefont {Fu}, \citenamefont {Cao}, \citenamefont {Zheng}, \citenamefont {Luo}, \citenamefont {Huang},\ and\ \citenamefont {Huang}}]{fu_first_2023}%
  \BibitemOpen
  \bibfield  {author} {\bibinfo {author} {\bibfnamefont {Z.}~\bibnamefont {Fu}}, \bibinfo {author} {\bibfnamefont {S.}~\bibnamefont {Cao}}, \bibinfo {author} {\bibfnamefont {H.}~\bibnamefont {Zheng}}, \bibinfo {author} {\bibfnamefont {J.}~\bibnamefont {Luo}}, \bibinfo {author} {\bibfnamefont {Q.}~\bibnamefont {Huang}},\ and\ \bibinfo {author} {\bibfnamefont {R.}~\bibnamefont {Huang}},\ }\bibfield  {title} {\bibinfo {title} {First {Demonstration} of {Hafnia}-based {Selector}-{Free} {FeRAM} with {High} {Disturb} {Immunity} through {Design} {Technology} {Co}-{Optimization}},\ }in\ \href {https://doi.org/10.1109/IEDM45741.2023.10413887} {\emph {\bibinfo {booktitle} {2023 {International} {Electron} {Devices} {Meeting} ({IEDM})}}}\ (\bibinfo  {publisher} {IEEE},\ \bibinfo {address} {San Francisco, CA, USA},\ \bibinfo {year} {2023})\ pp.\ \bibinfo {pages} {1--4}\BibitemShut {NoStop}%
\bibitem [{\citenamefont {Tagantsev}\ \emph {et~al.}(2010)\citenamefont {Tagantsev}, \citenamefont {Cross},\ and\ \citenamefont {Fousek}}]{tagantsev_domains_2010}%
  \BibitemOpen
  \bibfield  {author} {\bibinfo {author} {\bibfnamefont {A.~K.}\ \bibnamefont {Tagantsev}}, \bibinfo {author} {\bibfnamefont {L.~E.}\ \bibnamefont {Cross}},\ and\ \bibinfo {author} {\bibfnamefont {J.}~\bibnamefont {Fousek}},\ }\href {https://doi.org/10.1007/978-1-4419-1417-0} {\emph {\bibinfo {title} {Domains in {Ferroic} {Crystals} and {Thin} {Films}}}}\ (\bibinfo  {publisher} {Springer New York},\ \bibinfo {address} {New York, NY},\ \bibinfo {year} {2010})\BibitemShut {NoStop}%
\bibitem [{\citenamefont {Marton}\ \emph {et~al.}(2010)\citenamefont {Marton}, \citenamefont {Rychetsky},\ and\ \citenamefont {Hlinka}}]{Marton_domain_2010}%
  \BibitemOpen
  \bibfield  {author} {\bibinfo {author} {\bibfnamefont {P.}~\bibnamefont {Marton}}, \bibinfo {author} {\bibfnamefont {I.}~\bibnamefont {Rychetsky}},\ and\ \bibinfo {author} {\bibfnamefont {J.}~\bibnamefont {Hlinka}},\ }\bibfield  {title} {\bibinfo {title} {Domain walls of ferroelectric {BaTiO3} within the {Ginzburg}-{Landau}-{Devonshire} phenomenological model},\ }\href {https://doi.org/10.1103/PhysRevB.81.144125} {\bibfield  {journal} {\bibinfo  {journal} {Phys. Rev. B}\ }\textbf {\bibinfo {volume} {81}},\ \bibinfo {pages} {144125} (\bibinfo {year} {2010})}\BibitemShut {NoStop}%
\bibitem [{\citenamefont {Ishibashi}\ and\ \citenamefont {Takagi}(1971)}]{ishibashi_note_1971}%
  \BibitemOpen
  \bibfield  {author} {\bibinfo {author} {\bibfnamefont {Y.}~\bibnamefont {Ishibashi}}\ and\ \bibinfo {author} {\bibfnamefont {Y.}~\bibnamefont {Takagi}},\ }\bibfield  {title} {\bibinfo {title} {Note on {Ferroelectric} {Domain} {Switching}},\ }\href {https://doi.org/10.1143/JPSJ.31.506} {\bibfield  {journal} {\bibinfo  {journal} {J. Phys. Soc. Jpn.}\ }\textbf {\bibinfo {volume} {31}},\ \bibinfo {pages} {506} (\bibinfo {year} {1971})}\BibitemShut {NoStop}%
\bibitem [{\citenamefont {Tagantsev}\ \emph {et~al.}(2002)\citenamefont {Tagantsev}, \citenamefont {Stolichnov}, \citenamefont {Setter}, \citenamefont {Cross},\ and\ \citenamefont {Tsukada}}]{tagantsev_non-kolmogorov-avrami_2002}%
  \BibitemOpen
  \bibfield  {author} {\bibinfo {author} {\bibfnamefont {A.~K.}\ \bibnamefont {Tagantsev}}, \bibinfo {author} {\bibfnamefont {I.}~\bibnamefont {Stolichnov}}, \bibinfo {author} {\bibfnamefont {N.}~\bibnamefont {Setter}}, \bibinfo {author} {\bibfnamefont {J.~S.}\ \bibnamefont {Cross}},\ and\ \bibinfo {author} {\bibfnamefont {M.}~\bibnamefont {Tsukada}},\ }\bibfield  {title} {\bibinfo {title} {Non-{Kolmogorov}-{Avrami} switching kinetics in ferroelectric thin films},\ }\href {https://doi.org/10.1103/PhysRevB.66.214109} {\bibfield  {journal} {\bibinfo  {journal} {Phys. Rev. B}\ }\textbf {\bibinfo {volume} {66}},\ \bibinfo {pages} {214109} (\bibinfo {year} {2002})}\BibitemShut {NoStop}%
\bibitem [{\citenamefont {Merz}(1956)}]{merz_switching_1956}%
  \BibitemOpen
  \bibfield  {author} {\bibinfo {author} {\bibfnamefont {W.~J.}\ \bibnamefont {Merz}},\ }\bibfield  {title} {\bibinfo {title} {Switching {Time} in {Ferroelectric} {BaTiO3} and {Its} {Dependence} on {Crystal} {Thickness}},\ }\href {https://doi.org/10.1063/1.1722518} {\bibfield  {journal} {\bibinfo  {journal} {J. Appl. Phys.}\ }\textbf {\bibinfo {volume} {27}},\ \bibinfo {pages} {938} (\bibinfo {year} {1956})}\BibitemShut {NoStop}%
\bibitem [{\citenamefont {Pulvari}\ and\ \citenamefont {Kuebler}(1958)}]{pulvari_phenomenological_1958}%
  \BibitemOpen
  \bibfield  {author} {\bibinfo {author} {\bibfnamefont {C.~F.}\ \bibnamefont {Pulvari}}\ and\ \bibinfo {author} {\bibfnamefont {W.}~\bibnamefont {Kuebler}},\ }\bibfield  {title} {\bibinfo {title} {Phenomenological {Theory} of {Polarization} {Reversal} in {BaTiO3} {Single} {Crystals}},\ }\href {https://doi.org/10.1063/1.1723435} {\bibfield  {journal} {\bibinfo  {journal} {J. Appl. Phys.}\ }\textbf {\bibinfo {volume} {29}},\ \bibinfo {pages} {1315} (\bibinfo {year} {1958})}\BibitemShut {NoStop}%
\bibitem [{\citenamefont {Jo}\ \emph {et~al.}(2009)\citenamefont {Jo}, \citenamefont {Yang}, \citenamefont {Kim}, \citenamefont {Lee}, \citenamefont {Yoon}, \citenamefont {Park}, \citenamefont {Jo}, \citenamefont {Jung},\ and\ \citenamefont {Noh}}]{jo_nonlinear_2009}%
  \BibitemOpen
  \bibfield  {author} {\bibinfo {author} {\bibfnamefont {J.~Y.}\ \bibnamefont {Jo}}, \bibinfo {author} {\bibfnamefont {S.~M.}\ \bibnamefont {Yang}}, \bibinfo {author} {\bibfnamefont {T.~H.}\ \bibnamefont {Kim}}, \bibinfo {author} {\bibfnamefont {H.~N.}\ \bibnamefont {Lee}}, \bibinfo {author} {\bibfnamefont {J.-G.}\ \bibnamefont {Yoon}}, \bibinfo {author} {\bibfnamefont {S.}~\bibnamefont {Park}}, \bibinfo {author} {\bibfnamefont {Y.}~\bibnamefont {Jo}}, \bibinfo {author} {\bibfnamefont {M.~H.}\ \bibnamefont {Jung}},\ and\ \bibinfo {author} {\bibfnamefont {T.~W.}\ \bibnamefont {Noh}},\ }\bibfield  {title} {\bibinfo {title} {Nonlinear {Dynamics} of {Domain}-{Wall} {Propagation} in {Epitaxial} {Ferroelectric} {Thin} {Films}},\ }\href {https://doi.org/10.1103/PhysRevLett.102.045701} {\bibfield  {journal} {\bibinfo  {journal} {Phys. Rev. Lett.}\ }\textbf {\bibinfo {volume} {102}},\ \bibinfo {pages} {045701} (\bibinfo {year} {2009})}\BibitemShut {NoStop}%
\bibitem [{\citenamefont {Merz}(1954)}]{merz_domain_1954}%
  \BibitemOpen
  \bibfield  {author} {\bibinfo {author} {\bibfnamefont {W.~J.}\ \bibnamefont {Merz}},\ }\bibfield  {title} {\bibinfo {title} {Domain {Formation} and {Domain} {Wall} {Motions} in {Ferroelectric} {BaTiO} 3 {Single} {Crystals}},\ }\href {https://doi.org/10.1103/PhysRev.95.690} {\bibfield  {journal} {\bibinfo  {journal} {Phys. Rev.}\ }\textbf {\bibinfo {volume} {95}},\ \bibinfo {pages} {690} (\bibinfo {year} {1954})}\BibitemShut {NoStop}%
\bibitem [{\citenamefont {Lee}\ \emph {et~al.}(2021)\citenamefont {Lee}, \citenamefont {Lee}, \citenamefont {Yang}, \citenamefont {Park}, \citenamefont {Kim}, \citenamefont {Reddy}, \citenamefont {Materano}, \citenamefont {Mulaosmanovic}, \citenamefont {Mikolajick}, \citenamefont {Jones}, \citenamefont {Schroeder},\ and\ \citenamefont {Park}}]{lee_domains_2021}%
  \BibitemOpen
  \bibfield  {author} {\bibinfo {author} {\bibfnamefont {D.~H.}\ \bibnamefont {Lee}}, \bibinfo {author} {\bibfnamefont {Y.}~\bibnamefont {Lee}}, \bibinfo {author} {\bibfnamefont {K.}~\bibnamefont {Yang}}, \bibinfo {author} {\bibfnamefont {J.~Y.}\ \bibnamefont {Park}}, \bibinfo {author} {\bibfnamefont {S.~H.}\ \bibnamefont {Kim}}, \bibinfo {author} {\bibfnamefont {P.~R.~S.}\ \bibnamefont {Reddy}}, \bibinfo {author} {\bibfnamefont {M.}~\bibnamefont {Materano}}, \bibinfo {author} {\bibfnamefont {H.}~\bibnamefont {Mulaosmanovic}}, \bibinfo {author} {\bibfnamefont {T.}~\bibnamefont {Mikolajick}}, \bibinfo {author} {\bibfnamefont {J.~L.}\ \bibnamefont {Jones}}, \bibinfo {author} {\bibfnamefont {U.}~\bibnamefont {Schroeder}},\ and\ \bibinfo {author} {\bibfnamefont {M.~H.}\ \bibnamefont {Park}},\ }\bibfield  {title} {\bibinfo {title} {Domains and domain dynamics in fluorite-structured ferroelectrics},\ }\href {https://doi.org/10.1063/5.0047977} {\bibfield  {journal} {\bibinfo  {journal} {Appl. Phys. Rev}\ }\textbf
  {\bibinfo {volume} {8}},\ \bibinfo {pages} {021312} (\bibinfo {year} {2021})}\BibitemShut {NoStop}%
\bibitem [{\citenamefont {Alessandri}\ \emph {et~al.}(2018)\citenamefont {Alessandri}, \citenamefont {Pandey}, \citenamefont {Abusleme},\ and\ \citenamefont {Seabaugh}}]{alessandri_switching_2018}%
  \BibitemOpen
  \bibfield  {author} {\bibinfo {author} {\bibfnamefont {C.}~\bibnamefont {Alessandri}}, \bibinfo {author} {\bibfnamefont {P.}~\bibnamefont {Pandey}}, \bibinfo {author} {\bibfnamefont {A.}~\bibnamefont {Abusleme}},\ and\ \bibinfo {author} {\bibfnamefont {A.}~\bibnamefont {Seabaugh}},\ }\bibfield  {title} {\bibinfo {title} {Switching {Dynamics} of {Ferroelectric} {Zr}-{Doped} {HfO} $_{\textrm{2}}$},\ }\href {https://doi.org/10.1109/LED.2018.2872124} {\bibfield  {journal} {\bibinfo  {journal} {IEEE Electron Device Lett.}\ }\textbf {\bibinfo {volume} {39}},\ \bibinfo {pages} {1780} (\bibinfo {year} {2018})}\BibitemShut {NoStop}%
\bibitem [{\citenamefont {Liu}\ \emph {et~al.}(2016)\citenamefont {Liu}, \citenamefont {Grinberg},\ and\ \citenamefont {Rappe}}]{liu_intrinsic_2016}%
  \BibitemOpen
  \bibfield  {author} {\bibinfo {author} {\bibfnamefont {S.}~\bibnamefont {Liu}}, \bibinfo {author} {\bibfnamefont {I.}~\bibnamefont {Grinberg}},\ and\ \bibinfo {author} {\bibfnamefont {A.~M.}\ \bibnamefont {Rappe}},\ }\bibfield  {title} {\bibinfo {title} {Intrinsic ferroelectric switching from first principles},\ }\href {https://doi.org/10.1038/nature18286} {\bibfield  {journal} {\bibinfo  {journal} {Nature}\ }\textbf {\bibinfo {volume} {534}},\ \bibinfo {pages} {360} (\bibinfo {year} {2016})},\ \bibinfo {note} {publisher: Nature Publishing Group}\BibitemShut {NoStop}%
\bibitem [{\citenamefont {Su}\ and\ \citenamefont {Landis}(2007)}]{su_continuum_2007}%
  \BibitemOpen
  \bibfield  {author} {\bibinfo {author} {\bibfnamefont {Y.}~\bibnamefont {Su}}\ and\ \bibinfo {author} {\bibfnamefont {C.~M.}\ \bibnamefont {Landis}},\ }\bibfield  {title} {\bibinfo {title} {Continuum thermodynamics of ferroelectric domain evolution: {Theory}, finite element implementation, and application to domain wall pinning},\ }\href {https://doi.org/10.1016/j.jmps.2006.07.006} {\bibfield  {journal} {\bibinfo  {journal} {J. Mech. Phys. Solids}\ }\textbf {\bibinfo {volume} {55}},\ \bibinfo {pages} {280} (\bibinfo {year} {2007})}\BibitemShut {NoStop}%
\bibitem [{\citenamefont {Miller}\ and\ \citenamefont {Weinreich}(1960)}]{miller_mechanism_1960}%
  \BibitemOpen
  \bibfield  {author} {\bibinfo {author} {\bibfnamefont {R.~C.}\ \bibnamefont {Miller}}\ and\ \bibinfo {author} {\bibfnamefont {G.}~\bibnamefont {Weinreich}},\ }\bibfield  {title} {\bibinfo {title} {Mechanism for the sidewise motion of 180 domain walls in barium titanate},\ }\href {https://doi.org/10.1103/PhysRev.117.1460} {\bibfield  {journal} {\bibinfo  {journal} {Phys. Rev.}\ }\textbf {\bibinfo {volume} {117}},\ \bibinfo {pages} {1460} (\bibinfo {year} {1960})}\BibitemShut {NoStop}%
\bibitem [{\citenamefont {Shin}\ \emph {et~al.}(2007)\citenamefont {Shin}, \citenamefont {Grinberg}, \citenamefont {Chen},\ and\ \citenamefont {Rappe}}]{shin_nucleation_2007}%
  \BibitemOpen
  \bibfield  {author} {\bibinfo {author} {\bibfnamefont {Y.-H.}\ \bibnamefont {Shin}}, \bibinfo {author} {\bibfnamefont {I.}~\bibnamefont {Grinberg}}, \bibinfo {author} {\bibfnamefont {I.-W.}\ \bibnamefont {Chen}},\ and\ \bibinfo {author} {\bibfnamefont {A.~M.}\ \bibnamefont {Rappe}},\ }\bibfield  {title} {\bibinfo {title} {Nucleation and growth mechanism of ferroelectric domain-wall motion},\ }\href {https://doi.org/10.1038/nature06165} {\bibfield  {journal} {\bibinfo  {journal} {Nature}\ }\textbf {\bibinfo {volume} {449}},\ \bibinfo {pages} {881} (\bibinfo {year} {2007})}\BibitemShut {NoStop}%
\bibitem [{\citenamefont {Choe}\ \emph {et~al.}(2021)\citenamefont {Choe}, \citenamefont {Kim}, \citenamefont {Moon}, \citenamefont {Jo}, \citenamefont {Bae}, \citenamefont {Nam}, \citenamefont {Lee},\ and\ \citenamefont {Heo}}]{choe_unexpectedly_2021}%
  \BibitemOpen
  \bibfield  {author} {\bibinfo {author} {\bibfnamefont {D.-H.}\ \bibnamefont {Choe}}, \bibinfo {author} {\bibfnamefont {S.}~\bibnamefont {Kim}}, \bibinfo {author} {\bibfnamefont {T.}~\bibnamefont {Moon}}, \bibinfo {author} {\bibfnamefont {S.}~\bibnamefont {Jo}}, \bibinfo {author} {\bibfnamefont {H.}~\bibnamefont {Bae}}, \bibinfo {author} {\bibfnamefont {S.-G.}\ \bibnamefont {Nam}}, \bibinfo {author} {\bibfnamefont {Y.~S.}\ \bibnamefont {Lee}},\ and\ \bibinfo {author} {\bibfnamefont {J.}~\bibnamefont {Heo}},\ }\bibfield  {title} {\bibinfo {title} {Unexpectedly low barrier of ferroelectric switching in {HfO2} via topological domain walls},\ }\href {https://doi.org/10.1016/j.mattod.2021.07.022} {\bibfield  {journal} {\bibinfo  {journal} {Materials Today}\ }\textbf {\bibinfo {volume} {50}},\ \bibinfo {pages} {8} (\bibinfo {year} {2021})}\BibitemShut {NoStop}%
\bibitem [{\citenamefont {Lee}\ \emph {et~al.}(2020)\citenamefont {Lee}, \citenamefont {Lee}, \citenamefont {Lee}, \citenamefont {Jo}, \citenamefont {Yang}, \citenamefont {Kim}, \citenamefont {Chae}, \citenamefont {Waghmare},\ and\ \citenamefont {Lee}}]{lee_scale-free_2020}%
  \BibitemOpen
  \bibfield  {author} {\bibinfo {author} {\bibfnamefont {H.-J.}\ \bibnamefont {Lee}}, \bibinfo {author} {\bibfnamefont {M.}~\bibnamefont {Lee}}, \bibinfo {author} {\bibfnamefont {K.}~\bibnamefont {Lee}}, \bibinfo {author} {\bibfnamefont {J.}~\bibnamefont {Jo}}, \bibinfo {author} {\bibfnamefont {H.}~\bibnamefont {Yang}}, \bibinfo {author} {\bibfnamefont {Y.}~\bibnamefont {Kim}}, \bibinfo {author} {\bibfnamefont {S.~C.}\ \bibnamefont {Chae}}, \bibinfo {author} {\bibfnamefont {U.}~\bibnamefont {Waghmare}},\ and\ \bibinfo {author} {\bibfnamefont {J.~H.}\ \bibnamefont {Lee}},\ }\bibfield  {title} {\bibinfo {title} {Scale-free ferroelectricity induced by flat phonon bands in {HfO} $_{\textrm{2}}$},\ }\href {https://doi.org/10.1126/science.aba0067} {\bibfield  {journal} {\bibinfo  {journal} {Science}\ }\textbf {\bibinfo {volume} {369}},\ \bibinfo {pages} {1343} (\bibinfo {year} {2020})}\BibitemShut {NoStop}%
\bibitem [{\citenamefont {Ding}\ \emph {et~al.}(2020)\citenamefont {Ding}, \citenamefont {Zhang}, \citenamefont {Tao}, \citenamefont {Yang},\ and\ \citenamefont {Zhou}}]{ding_atomic-scale_2020}%
  \BibitemOpen
  \bibfield  {author} {\bibinfo {author} {\bibfnamefont {W.}~\bibnamefont {Ding}}, \bibinfo {author} {\bibfnamefont {Y.}~\bibnamefont {Zhang}}, \bibinfo {author} {\bibfnamefont {L.}~\bibnamefont {Tao}}, \bibinfo {author} {\bibfnamefont {Q.}~\bibnamefont {Yang}},\ and\ \bibinfo {author} {\bibfnamefont {Y.}~\bibnamefont {Zhou}},\ }\bibfield  {title} {\bibinfo {title} {The atomic-scale domain wall structure and motion in {HfO2}-based ferroelectrics: {A} first-principle study},\ }\href {https://doi.org/10.1016/j.actamat.2020.07.012} {\bibfield  {journal} {\bibinfo  {journal} {Acta Materialia}\ }\textbf {\bibinfo {volume} {196}},\ \bibinfo {pages} {556} (\bibinfo {year} {2020})}\BibitemShut {NoStop}%
\bibitem [{\citenamefont {Cahn}(1960)}]{cahn_theory_1960}%
  \BibitemOpen
  \bibfield  {author} {\bibinfo {author} {\bibfnamefont {J.~W.}\ \bibnamefont {Cahn}},\ }\bibfield  {title} {\bibinfo {title} {Theory of crystal growth and interface motion in crystalline materials},\ }\href {https://doi.org/10.1016/0001-6160(60)90110-3} {\bibfield  {journal} {\bibinfo  {journal} {Acta Metallurgica}\ }\textbf {\bibinfo {volume} {8}},\ \bibinfo {pages} {554} (\bibinfo {year} {1960})}\BibitemShut {NoStop}%
\bibitem [{\citenamefont {Wang}\ \emph {et~al.}(2019)\citenamefont {Wang}, \citenamefont {Wang},\ and\ \citenamefont {Chen}}]{wang_understanding_2019}%
  \BibitemOpen
  \bibfield  {author} {\bibinfo {author} {\bibfnamefont {J.-J.}\ \bibnamefont {Wang}}, \bibinfo {author} {\bibfnamefont {B.}~\bibnamefont {Wang}},\ and\ \bibinfo {author} {\bibfnamefont {L.-Q.}\ \bibnamefont {Chen}},\ }\bibfield  {title} {\bibinfo {title} {Understanding, {Predicting}, and {Designing} {Ferroelectric} {Domain} {Structures} and {Switching} {Guided} by the {Phase}-{Field} {Method}},\ }\href {https://doi.org/10.1146/annurev-matsci-070218-121843} {\bibfield  {journal} {\bibinfo  {journal} {Annu. Rev. Mater. Res.}\ }\textbf {\bibinfo {volume} {49}},\ \bibinfo {pages} {127} (\bibinfo {year} {2019})}\BibitemShut {NoStop}%
\bibitem [{\citenamefont {Hlinka}\ and\ \citenamefont {Márton}(2006)}]{hlinka_phenomenological_2006}%
  \BibitemOpen
  \bibfield  {author} {\bibinfo {author} {\bibfnamefont {J.}~\bibnamefont {Hlinka}}\ and\ \bibinfo {author} {\bibfnamefont {P.}~\bibnamefont {Márton}},\ }\bibfield  {title} {\bibinfo {title} {Phenomenological model of a 90° domain wall in {BaTiO3}-type ferroelectrics},\ }\href@noop {} {\bibfield  {journal} {\bibinfo  {journal} {Phys. Rev. B}\ }\textbf {\bibinfo {volume} {74}} (\bibinfo {year} {2006})}\BibitemShut {NoStop}%
\bibitem [{\citenamefont {Xiao}\ \emph {et~al.}(2005)\citenamefont {Xiao}, \citenamefont {Shenoy},\ and\ \citenamefont {Bhattacharya}}]{xiao_depletion_2005}%
  \BibitemOpen
  \bibfield  {author} {\bibinfo {author} {\bibfnamefont {Y.}~\bibnamefont {Xiao}}, \bibinfo {author} {\bibfnamefont {V.~B.}\ \bibnamefont {Shenoy}},\ and\ \bibinfo {author} {\bibfnamefont {K.}~\bibnamefont {Bhattacharya}},\ }\bibfield  {title} {\bibinfo {title} {Depletion {Layers} and {Domain} {Walls} in {Semiconducting} {Ferroelectric} {Thin} {Films}},\ }\href {https://doi.org/10.1103/PhysRevLett.95.247603} {\bibfield  {journal} {\bibinfo  {journal} {Phys. Rev. Lett.}\ }\textbf {\bibinfo {volume} {95}},\ \bibinfo {pages} {247603} (\bibinfo {year} {2005})}\BibitemShut {NoStop}%
\bibitem [{Note1()}]{Note1}%
  \BibitemOpen
  \bibinfo {note} {The analytical solution is $P_y = -P_0 + 2P_0 \tanh (\protect \frac {x-x_{DW}}{\delta })$, $P_x = 0$}\BibitemShut {NoStop}%
\bibitem [{\citenamefont {Collins}\ \emph {et~al.}(1979)\citenamefont {Collins}, \citenamefont {Blumen}, \citenamefont {Currie},\ and\ \citenamefont {Ross}}]{collins_dynamics_1979}%
  \BibitemOpen
  \bibfield  {author} {\bibinfo {author} {\bibfnamefont {M.~A.}\ \bibnamefont {Collins}}, \bibinfo {author} {\bibfnamefont {A.}~\bibnamefont {Blumen}}, \bibinfo {author} {\bibfnamefont {J.~F.}\ \bibnamefont {Currie}},\ and\ \bibinfo {author} {\bibfnamefont {J.}~\bibnamefont {Ross}},\ }\bibfield  {title} {\bibinfo {title} {Dynamics of domain walls in ferrodistortive materials. {I}. {Theory}},\ }\href {https://doi.org/10.1103/PhysRevB.19.3630} {\bibfield  {journal} {\bibinfo  {journal} {Phys. Rev. B}\ }\textbf {\bibinfo {volume} {19}},\ \bibinfo {pages} {3630} (\bibinfo {year} {1979})}\BibitemShut {NoStop}%
\bibitem [{Note2()}]{Note2}%
  \BibitemOpen
  \bibinfo {note} {The analytical solution is $P_y = -\protect \frac {P_0}{\protect \sqrt {2}} + \protect \sqrt {2}P_0 \tanh (\protect \frac {x-x_{DW}}{\delta }\protect \sqrt {\protect \frac {2\kappa }{\kappa ' + \kappa }})$ , $P_x = P_0/\protect \sqrt {2}$}\BibitemShut {NoStop}%
\bibitem [{Note3()}]{Note3}%
  \BibitemOpen
  \bibinfo {note} {The key step of this derivation is $\DOTSB \sum@ \slimits@ _n f_{grad,n} = \protect \frac {\kappa }{2 h_l^2} \DOTSB \sum@ \slimits@ _n (P_{n+1} - P_n)^2 = \protect \frac {\kappa }{2 h_l^2} \DOTSB \sum@ \slimits@ _n P_n(2P_n - P_{n-1} - P_{n+1}) = \protect \frac {1}{2} \DOTSB \sum@ \slimits@ _n P_n (\protect \frac {\partial f_{landau}}{\partial P})_n$}\BibitemShut {NoStop}%
\bibitem [{Note4()}]{Note4}%
  \BibitemOpen
  \bibinfo {note} {The key step of this derivation is to use $P_n - P_{n-1} \approx P_0(1-(\protect \frac {P_n}{P_0})^2) \protect \frac {h_l}{\delta }$ (using the Taylor expansion of the tanh function) to calculate the gradient energy $\protect \frac {\kappa }{2 h_l^2} \DOTSB \sum@ \slimits@ _n (P_{n+1} - P_n)^2$}\BibitemShut {NoStop}%
\bibitem [{\citenamefont {Meyer}\ and\ \citenamefont {Vanderbilt}(2002)}]{meyer_ab_2002}%
  \BibitemOpen
  \bibfield  {author} {\bibinfo {author} {\bibfnamefont {B.}~\bibnamefont {Meyer}}\ and\ \bibinfo {author} {\bibfnamefont {D.}~\bibnamefont {Vanderbilt}},\ }\bibfield  {title} {\bibinfo {title} {\textit{{Ab} initio} study of ferroelectric domain walls in {PbTiO} 3},\ }\href {https://doi.org/10.1103/PhysRevB.65.104111} {\bibfield  {journal} {\bibinfo  {journal} {Phys. Rev. B}\ }\textbf {\bibinfo {volume} {65}},\ \bibinfo {pages} {104111} (\bibinfo {year} {2002})}\BibitemShut {NoStop}%
\bibitem [{\citenamefont {Cao}\ and\ \citenamefont {Cross}(1991)}]{cao_theory_1991}%
  \BibitemOpen
  \bibfield  {author} {\bibinfo {author} {\bibfnamefont {W.}~\bibnamefont {Cao}}\ and\ \bibinfo {author} {\bibfnamefont {L.~E.}\ \bibnamefont {Cross}},\ }\bibfield  {title} {\bibinfo {title} {Theory of tetragonal twin structures in ferroelectric perovskites with a first-order phase transition},\ }\href@noop {} {\bibfield  {journal} {\bibinfo  {journal} {Phys. Rev. B}\ }\textbf {\bibinfo {volume} {44}} (\bibinfo {year} {1991})}\BibitemShut {NoStop}%
\bibitem [{\citenamefont {Li}\ \emph {et~al.}(2002)\citenamefont {Li}, \citenamefont {Hu}, \citenamefont {Liu},\ and\ \citenamefont {Chen}}]{li_effect_2002}%
  \BibitemOpen
  \bibfield  {author} {\bibinfo {author} {\bibfnamefont {Y.}~\bibnamefont {Li}}, \bibinfo {author} {\bibfnamefont {S.}~\bibnamefont {Hu}}, \bibinfo {author} {\bibfnamefont {Z.}~\bibnamefont {Liu}},\ and\ \bibinfo {author} {\bibfnamefont {L.}~\bibnamefont {Chen}},\ }\bibfield  {title} {\bibinfo {title} {Effect of substrate constraint on the stability and evolution of ferroelectric domain structures in thin films},\ }\href {https://doi.org/10.1016/S1359-6454(01)00360-3} {\bibfield  {journal} {\bibinfo  {journal} {Acta Materialia}\ }\textbf {\bibinfo {volume} {50}},\ \bibinfo {pages} {395} (\bibinfo {year} {2002})}\BibitemShut {NoStop}%
\bibitem [{\citenamefont {Zhang}\ and\ \citenamefont {Bhattacharya}(2005)}]{zhang_computational_2005}%
  \BibitemOpen
  \bibfield  {author} {\bibinfo {author} {\bibfnamefont {W.}~\bibnamefont {Zhang}}\ and\ \bibinfo {author} {\bibfnamefont {K.}~\bibnamefont {Bhattacharya}},\ }\bibfield  {title} {\bibinfo {title} {A computational model of ferroelectric domains. {Part} {I}: model formulation and domain switching},\ }\href@noop {} {\bibfield  {journal} {\bibinfo  {journal} {Acta Materialia}\ } (\bibinfo {year} {2005})}\BibitemShut {NoStop}%
\bibitem [{\citenamefont {Chandra}\ and\ \citenamefont {Littlewood}(2007)}]{chandra_landau_2007}%
  \BibitemOpen
  \bibfield  {author} {\bibinfo {author} {\bibfnamefont {P.}~\bibnamefont {Chandra}}\ and\ \bibinfo {author} {\bibfnamefont {P.~B.}\ \bibnamefont {Littlewood}},\ }\bibfield  {title} {\bibinfo {title} {A {Landau} {Primer} for {Ferroelectrics}},\ }in\ \href {https://doi.org/10.1007/978-3-540-34591-6_3} {\emph {\bibinfo {booktitle} {Physics of {Ferroelectrics}}}},\ Vol.\ \bibinfo {volume} {105}\ (\bibinfo  {publisher} {Springer Berlin Heidelberg},\ \bibinfo {address} {Berlin, Heidelberg},\ \bibinfo {year} {2007})\ pp.\ \bibinfo {pages} {69--116},\ \bibinfo {note} {series Title: Topics in Applied Physics}\BibitemShut {NoStop}%
\end{thebibliography}%

\end{document}